# GSC 02626-00896: an RR Lyrae star with a ceasing Blazhko effect and three new variables in the field.

Rainer Gröbel [1, 2]

1) DE- 90542 Eckental, Germany rainer.groebel@web.de
2) Bundesdeutsche Arbeitsgemeinschaft für Veränderliche Sterne e.V. (BAV)
Munsterdamm 90, D-12169 Berlin, Germany

BAV Mitteilungen Nr. 230

**Abstract:** By analysis of Superwasp data, Srdoc and Bernhard (2012) have shown recently that GSC 02626-00896 (18:09:30.34 + 32:45:13.52) is an RRc type variable showing indications of Blazhko effect. A night by night reanalysis of the data shows that in fact a 26 d secondary period is present in the 2004 observations. However, no trace of this can be found in the 2007 and 2008 data. CCD observations of the star on 19 nights in 2012 yielded a strictly repeating light curve with a broad maximum similar to those derived from the second and third SWASP data group. A slightly revised ephemeris Max.(HJD) = 2456069.3919(43) + 0.32272138(60) d x E could be derived.

The CCD observations also revealed three short period eclipsing variables in the field of GSC 02626-00896. The 15.16 R mag. GSC2.3 star N25F001920 (18:09:24.282 +32:44:51.72) shows 0.15 and 0.1 mag. shallow eclipses with the ephemeris Max.(HJD) = 2456064.5155(13) + 0.389228(18) d x E. The 15.39 R mag. GSC 2.3 N25D013720 (18:09:41.141 +32:33:46.99) is a W UMa type star with the ephemeris Min.(HJD) = 2456072.4354(4) + 0.393563(7) d x E and 0.45 mag. amplitude. The 16.13 R mag. GSC 2.3 N25D015132 (18:09:01.767 +32:36:53.59) is also a W UMa type star with the ephemeris Min.(HJD) = 2456072.4909(2) + 0.287800(3) d x E with an amplitude of 0.75 and 0.5 mag respectively.

**Simbad objects:** GSC1 02626-00896  GSC2.3 N25F001920  GSC2.3 N25D013720  GSC2.3 N25D015132

**1) GSC 02626-00896 (18:09:30.34 + 32:45:13.52),** 2MASS 18093032+3245136, 1SWASP J180930.33 +324513.8, USNO-B1.0 1227-0378420:

The variability of the star was discovered by G.Srdoc and K. Bernhard (Srdoc and Bernhard, 2012) by analysis of SuperWASP data (Butters et al., 2010). It was classified as an RRc star with the ephemeris:

$$\text{HJD (Max.)} = 2453160.713(3) + 0.3227214(2) \times E \qquad (1)$$

From the variability of the light curve (LC), it was presumed that Blazhko effect is present. For the present analysis SWASP data was reanalysed and the star was included in the observation program.

**SWASP data analysis:**

The SWASP database gives 16946 measurements for this star. They are grouped into three dense observation periods lasting from April 29 to October 01 2004, from July 15 to August 23 2007 and from April 18 to August 09 2008 (fig.1).

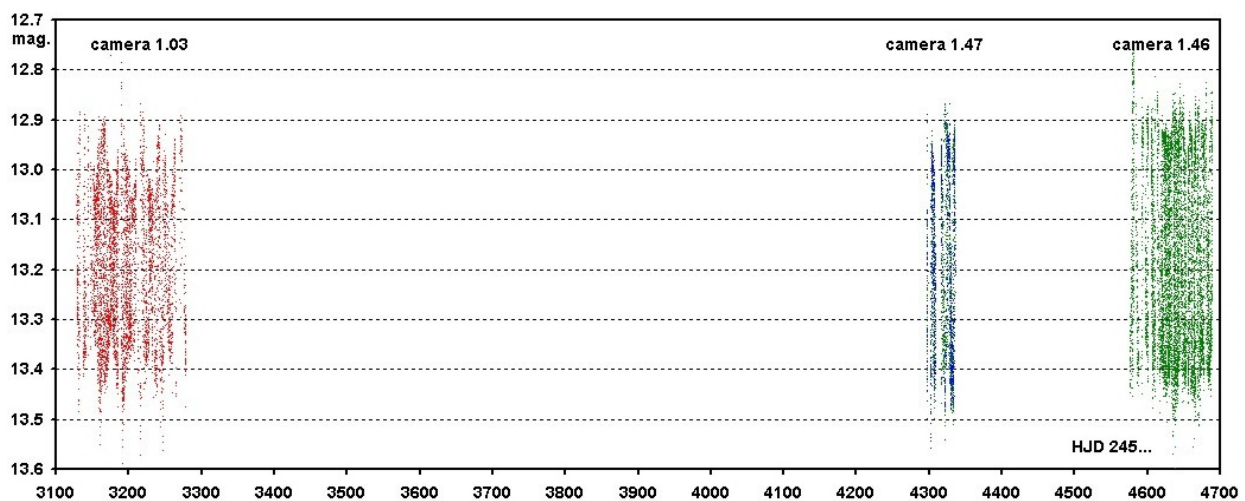

**Figure 1:** The presently available SWASP data on GSC 02626-00896.

The data was first sorted by the camera identification number (ID). Only the TAMFlux 2 magnitudes were used for they are corrected for the differing sensitivity of the involved cameras. Furthermore, data with a TAMFlux 2 error exceeding +/- 0.1 mag. was discarded. The data was displayed night by night and the eventually present maxima or minima in the LC were isolated and redisplayed at closer scale. A sixth order polynomial was used to determine the time of extrema, but only if the extrema were sufficiently covered. In most cases, the error in time determination was estimated to be +/-0.001 d.

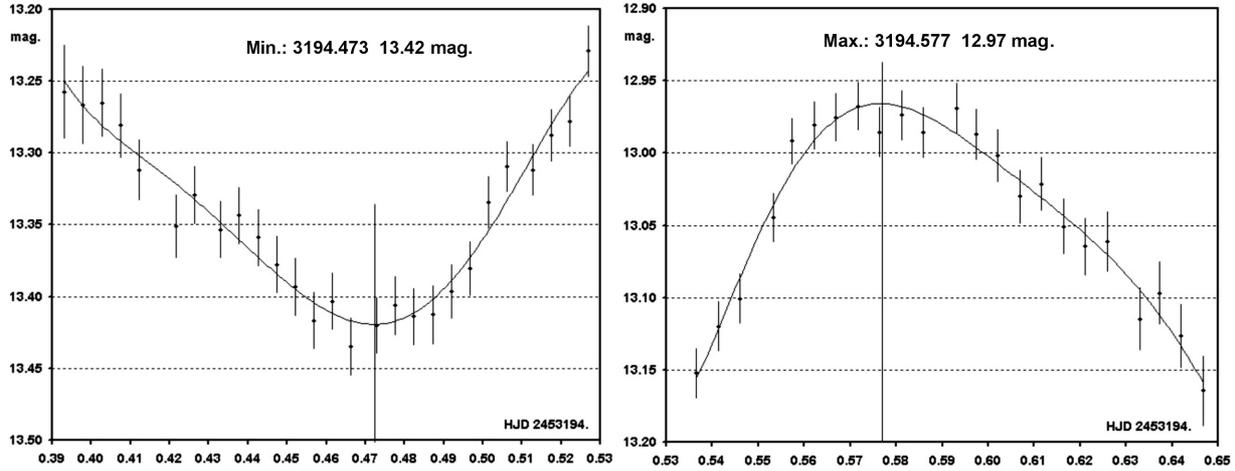

**Figure 2:** The minimum and maximum determination in the SWASP LC for the 7.7.2004 observations.

**The first SWASP data group:**

In fig. 3 the 2004 observations are shown at greater scale. A rhythmic variation of the amplitude of the variable between 0.3 to 0.6 mag. can be seen whose period was estimated to be somewhat shorter than 30 d. This was a first indication of a pronounced and relatively rapid Blazhko effect.

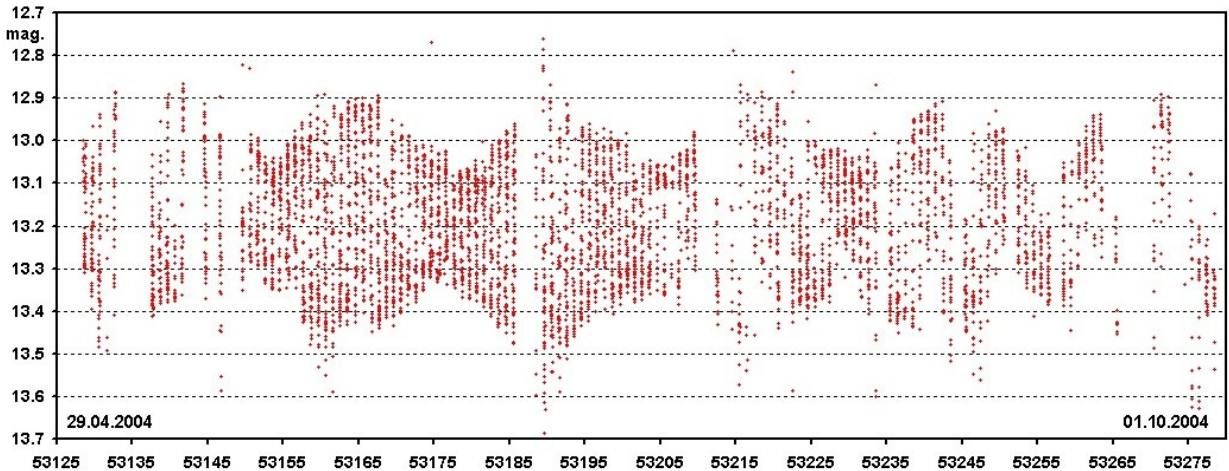

**Figure 3:** The first observation period, spanning around five months.

From data of this observation period, the times and magnitudes of 64 maxima (max.) and 59 minima (min.) (Table 1) could be determined. In the (O-C) diagram (fig. 4), the max. timing varies between one hour later and half an hour earlier against the calculations with a mean period of 0.3227 d. An approximate secondary period of 26 d was derived and could be followed over five cycles. It seems that the min. doesn´t follow the cycle. This may be due to variations in the steepness of the rising branch of the LC. An investigation will require LCs with both extrema in one night, but this is only rarely found in SWASP data.

In contrast, the magnitude variations diagram of the extrema (fig. 5) shows that if the max. are bright, the min. are faint and vice versa. This alternation follows the Blazhko cycle and explains the above noticed strong variations in the amplitude of the LC.

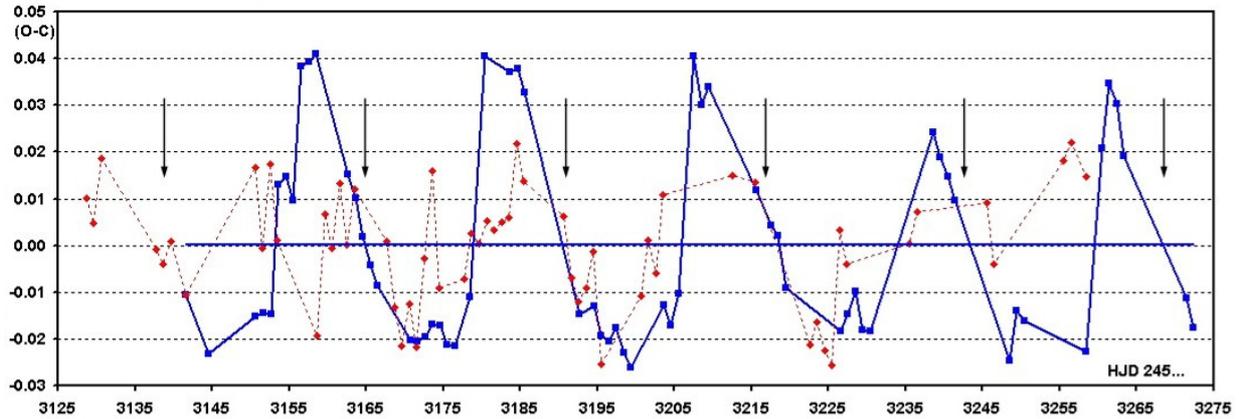

**Figure 4:** The deviations in time of the max. (blue) and min. (red) against the mean period. The 26d Blazhko period is marked with arrows.

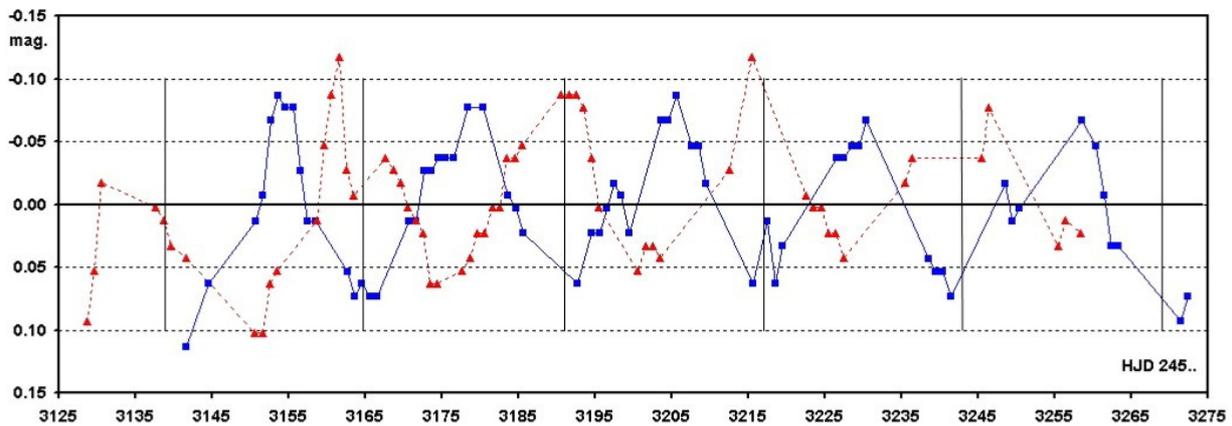

**Figure 5:** The magnitude variations of the max. (blue) and min. (red) against their respective mean magnitude.

| Max. JD hel. | Epoch | (O-C) | Min. JD hel. | Epoch | (O-C) |
|---|---|---|---|---|---|
| 2453141.650 | -9072 | -0.014 | 2453141.540 | -9057 | -0.008 |
| 2453144.542 | -9063 | -0.026 | 2453150.603 | -9029 | 0.019 |
| 2453150.682 | -9044 | -0.018 | 2453151.554 | -9026 | 0.002 |
| 2453151.651 | -9041 | -0.017 | 2453152.540 | -9023 | 0.019 |
| 2453152.619 | -9038 | -0.017 | 2453153.492 | -9020 | 0.003 |
| 2453153.615 | -9035 | 0.011 | 2453158.635 | -9004 | -0.017 |
| 2453154.585 | -9032 | 0.013 | 2453159.629 | -9001 | 0.009 |
| 2453155.548 | -9029 | 0.007 | 2453160.590 | -8998 | 0.001 |
| 2453156.545 | -9026 | 0.036 | 2453161.572 | -8995 | 0.015 |
| 2453157.514 | -9023 | 0.037 | 2453162.527 | -8992 | 0.002 |
| 2453158.484 | -9020 | 0.039 | 2453163.507 | -8989 | 0.014 |
| 2453162.654 | -9007 | 0.014 | 2453167.691 | -8976 | 0.003 |
| 2453163.617 | -9004 | 0.008 | 2453168.645 | -8973 | -0.012 |
| 2453164.577 | -9001 | 0.000 | 2453169.605 | -8970 | -0.020 |
| 2453165.539 | -8998 | -0.006 | 2453170.582 | -8967 | -0.011 |
| 2453166.503 | -8995 | -0.010 | 2453171.541 | -8964 | -0.020 |
| 2453170.687 | -8982 | -0.021 | 2453172.528 | -8961 | -0.001 |
| 2453171.655 | -8979 | -0.022 | 2453173.515 | -8958 | 0.018 |
| 2453172.624 | -8976 | -0.021 | 2453174.458 | -8955 | -0.008 |
| 2453173.595 | -8973 | -0.018 | 2453177.687 | -8945 | -0.006 |

**Table 1:** The SWASP max. and min. over the first Blazhko period. The (O-C) were calculated with ephemeris (2) and (3) respectively. The timings of all extrema are given in appendix 1.

Another way to illustrate the behaviour of the Blazhko cycle is shown in fig. 6. Starting from right counterclockwise, the max. are progressively getting brighter and coming in earlier until they reach their greatest advance in time at a mean magnitude around 13 mag.. Then they get fainter with increasing retardation in time. In Le Borgne et al., 2010, a great diversity of cycle shapes is shown. This seems to be specific for a given Blazhko RR Lyrae star.

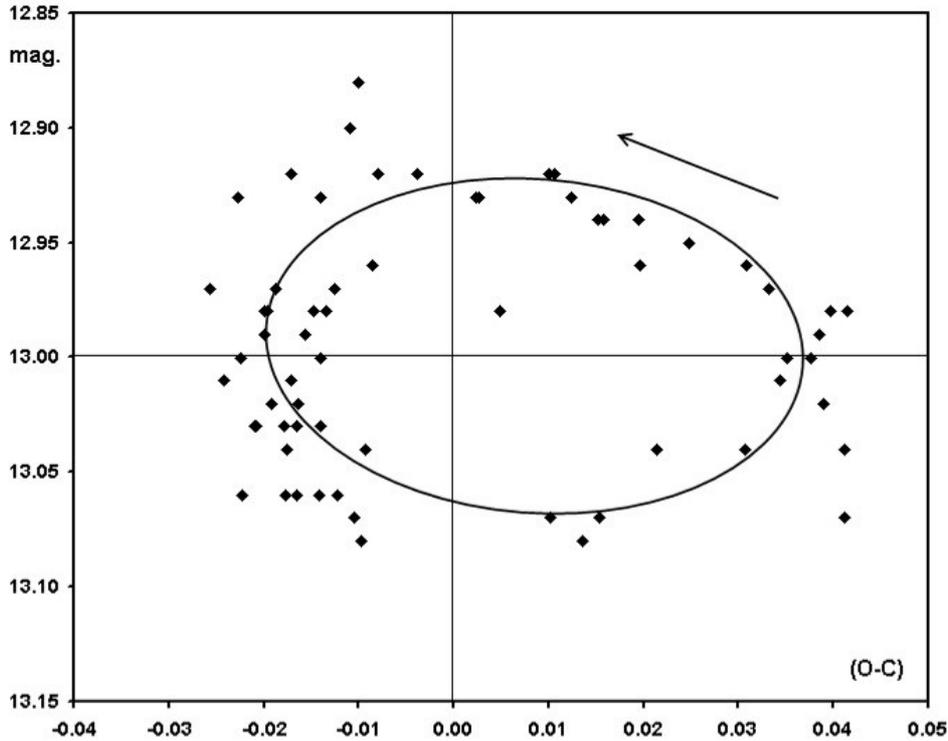

**Figure 6:** In 2004 the max. followed a 26 d cycle counterclockwise.

**The second and third SWASP data group:**

The secondary period seems to be characteristic for a given Blazhko RR Lyrae star. Cycles fom a few to several hundred days are known with a clustering between 20 and 40 d. Thus it was surprising to find no trace of a secondary period during the processing of the second data group (Fig. 7, middle). The overall luminosity increased about 0.3 mag. and the star seems to pulsate with its mean period displaying only slight variations in the shape of the LC, at the limit of the data scatter. Somewhere between late 2004 and mid 2007 a major change in the pulsation mode must have taken place.

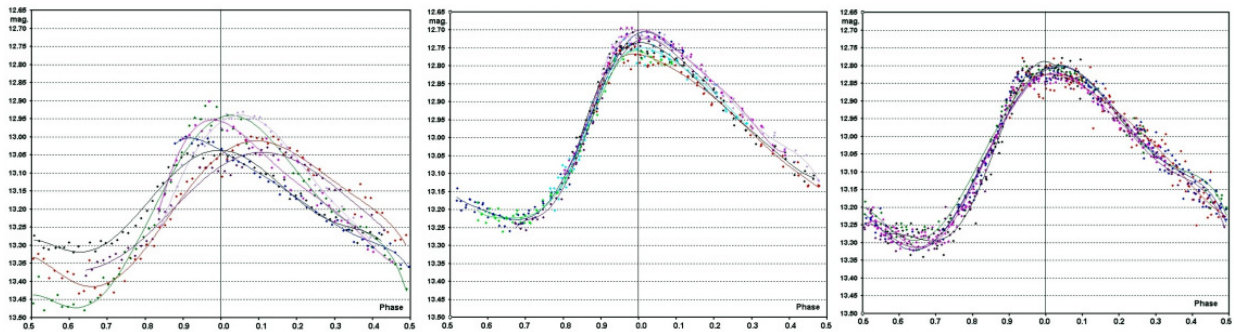

**Figure 7:** A sample of LCs from the three SWASP observation periods.

In the third data group the overall luminosity decreases by about 0.1 mag., but the LC shows a similar shape with only slight variations. The 2012 observations (Fig. 9) show an analogous shape and amplitude of the LC with rounded max. and only slight variations in their height. There seems to be no variation of the mean period; all max. fit well in the (O-C) diagram in fig. 10 on the base of ephemeris (2).

It will be interesting to investigate when and how the transition has taken place, but observations of this formerly unknown variable are probably not available. A search for data from other robotic telescopes yielded only a data set from the CSDR2 (The Catalina Sky Survey). The observations span from April 18 2005 to June 04 2006, eight months after the first SWASP serie. The few measurements at least demonstrate that the LC also shows great variations in amplitude; therefore the Blazhko cycle may still be present at that time.

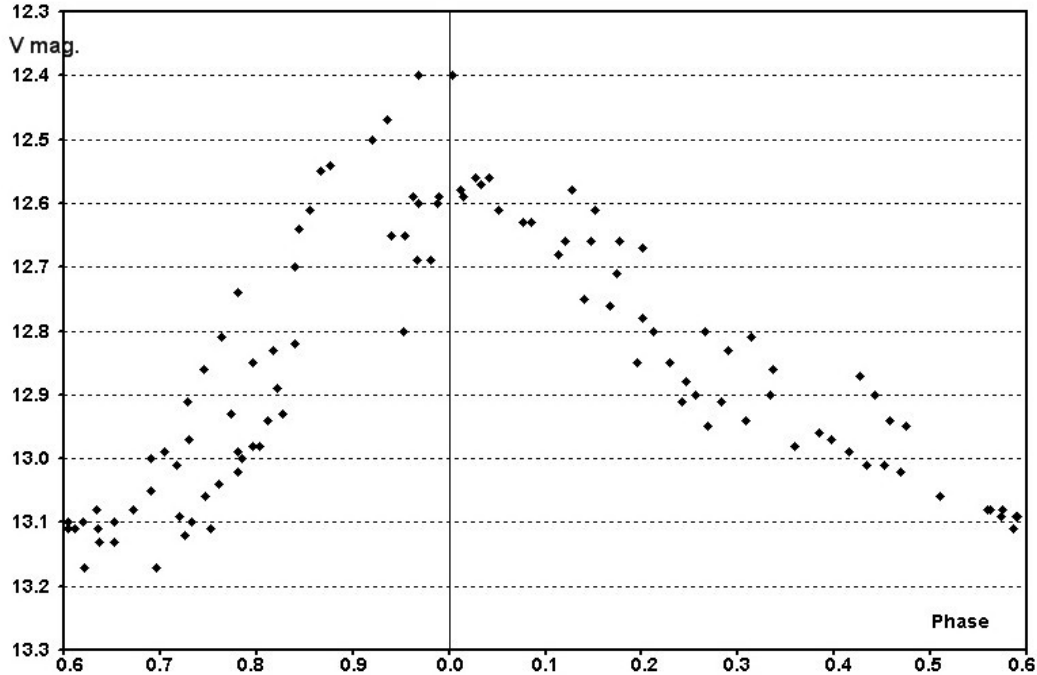

**Figure 8:** The measurements of the CRTS robotic telescope reduced with ephemeris (2).

By comparing the LCs in fig. 7, a problem of the SWASP data appears. In the measurements different cameras with unequal sensitive CCD chips are involved. The uncorrected Flux2 magnitudes show variations up to +/- 0.2 mag. from one camera to another at the same observation time. To align the cameras on a common magnitude scale, the SWASP group developed a correction program (Collier Cameron et al., 2006). Beside the raw data, a corrected TAMFlux 2 value and the corresponding TAMFlux2 error is given. Inspection of the data shows that the magnitude values are now much closer, but the authors warned that their solution may not be infaillible.

A simple way to control the validity of the correction is to take some of the comparison stars used in the 2012 measurements (Fig. 11) and check the constancy of their magnitude values from one camera to another. The mean magnitudes for the involved cameras fit well with only differences of a few millimags. and standard deviations about +/- 0.02 mag. (Table 2). The deviation of check star 1 with camera 1.03 could not be explained. The 2012 measurements showed no variations against the other reference stars exceeding the scatter.

| star | Camera ID | 1.03 | | 1.46 I | | 1.46 II | | 1.47 | |
|---|---|---|---|---|---|---|---|---|---|
| GSC1 | SWASP | Tmag. | +/- | Tmag. | +/- | Tmag. | +/- | Tmag. | +/- |
| 2626 0267 | Comp. | 12.841 | 0.018 | 12.840 | 0.014 | 12.841 | 0.019 | 12.839 | 0.018 |
| 2626 1106 | Chk 1 | 13.100 | 0.028 | 13.139 | 0.019 | 13.141 | 0.025 | 13.139 | 0.025 |
| 2626 0226 | Chk 2 | 12.498 | 0.015 | 12.497 | 0.012 | 12.498 | 0.014 | 12.496 | 0.014 |
| 2626 0643 | Chk 3 | 11.535 | 0.018 | 11.534 | 0.012 | 11.535 | 0.014 | 11.533 | 0.015 |
| 2626 1155 | Chk 4 | 11.662 | 0.024 | 11.660 | 0.020 | 11.662 | 0.022 | 11.658 | 0.022 |

**Table 2:** Comparison stars magnitudes for different cameras.

**The 2012 measurements:**

The measurements were performed with a 10" SCT in a semi-automated mode and a SBIG ST8XME camera with 94s exposition time in the 2x2 binning mode without filter to increase the S/N ratio. The observation period lasted from May 22 to July 29 2012 under mostly good sky conditions. 19 series yielding a total of 2641 measurements could be won. The reductions were performed with the Muniwin (Motl, D., 2012) reduction program. Depending on weather conditions, a 3.8, 5 or 7 pixel radius diaphragm corresponding to a 6, 7.5 or 10.5" aperture radius was used. Twilight sky-flat images were used for flatfield corrections.

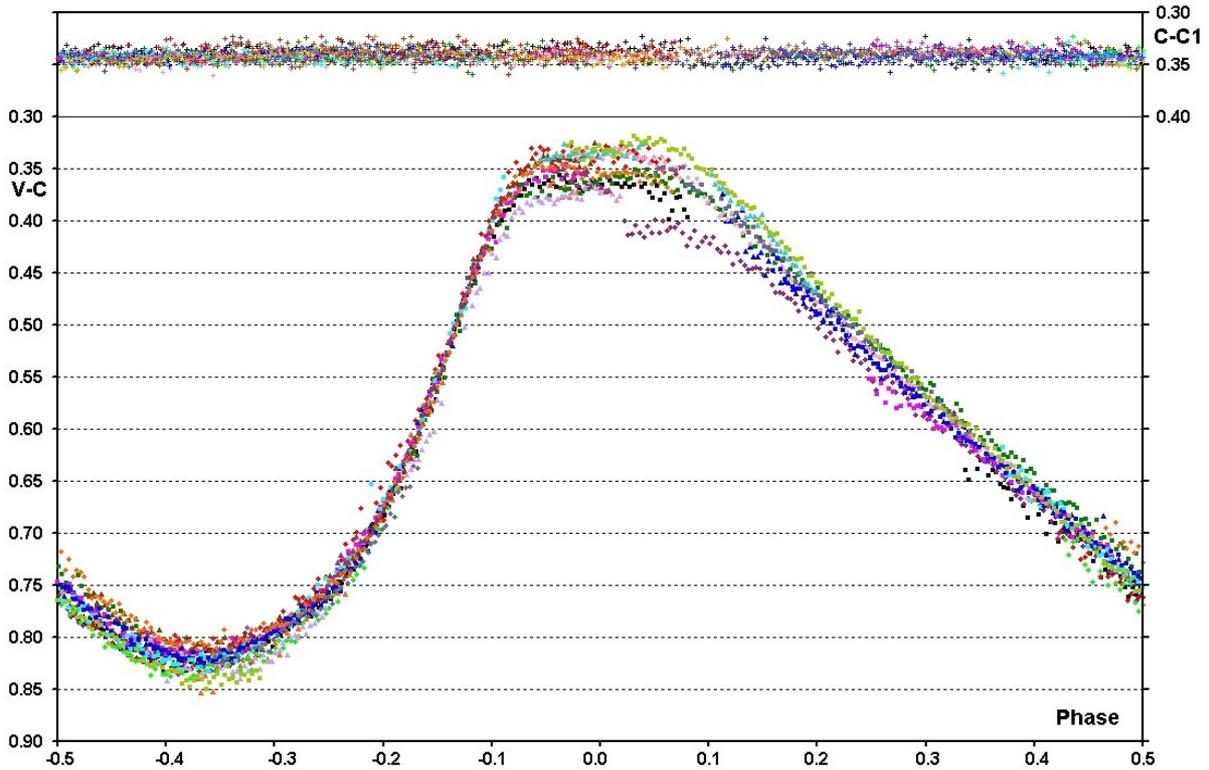

**Figure 9:** The 2012 measurements reduced with ephemeris (2).

The LC in fig. 9 shows an overall constant shape with only some slight variations in the height of the max. Three max. and six min. could be determined and with the SWASP extrema, the ephemeris

$$\text{HJD (Max.)} = 2456069.3919(43) + 0.32272138(60) * E \quad (2)$$

was derived. The deviations of the min. times were calculated with the ephemeris

$$\text{HJD (Min.)} = 2456064.4372(16) + 0.32272156(24) * E \quad (3)$$

The timings of all extrema yielded the (O-C) diagram in fig.10 and are tabulated in appendix 1.

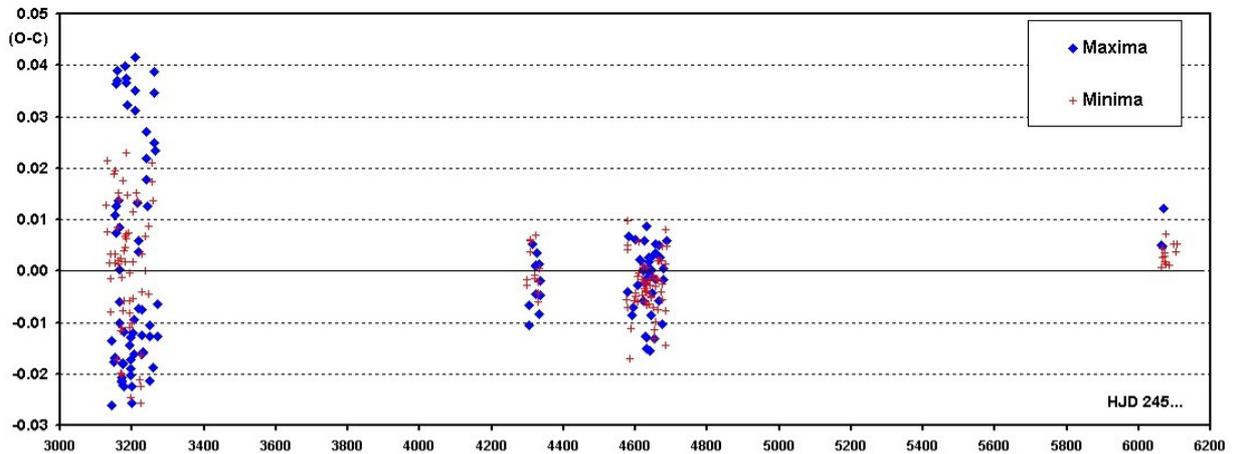

**Figure 10:** (O-C) diagram with the SWASP and the 2012 extrema.

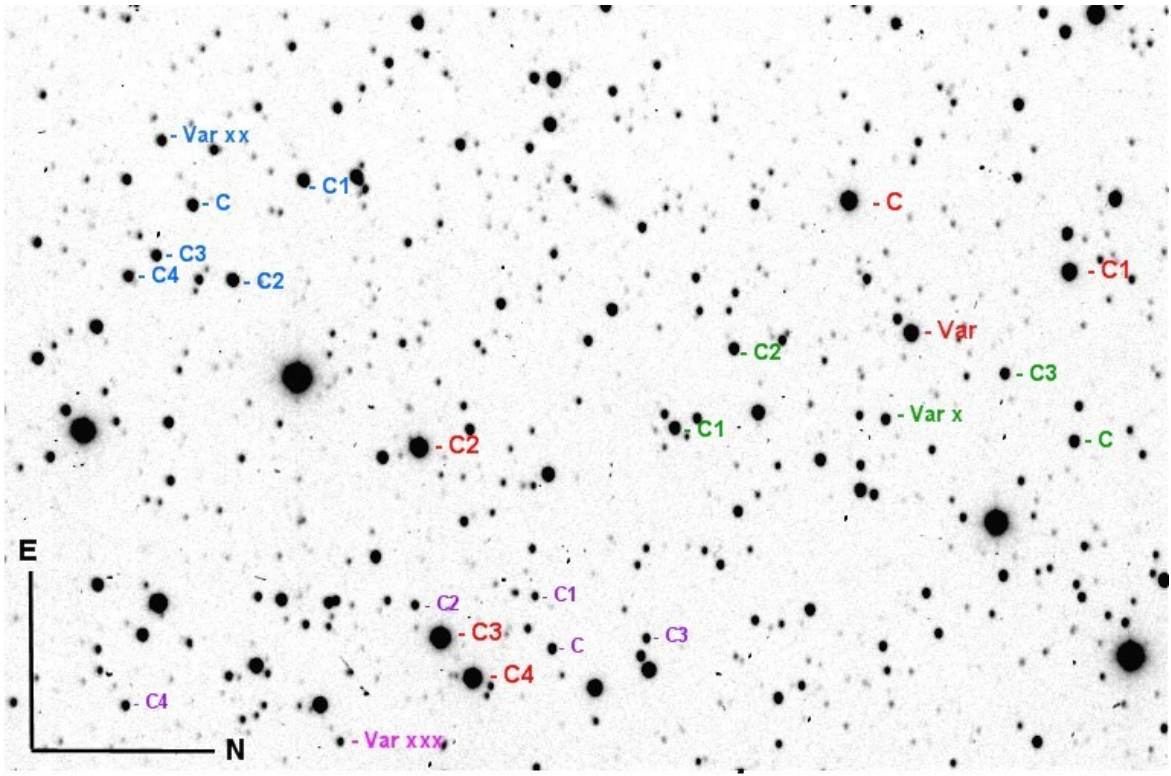

**Fig. 11:** The 20´ x 13´ field of the camera with the RR lyrae star Var, the variables Vx, Vxx, Vxxx and their comparison stars.

**2) Var. X (18:09:24.282 +32:44:51.72)**, GSC2.3 N25F001920, 2MASS 18092430+3244515:

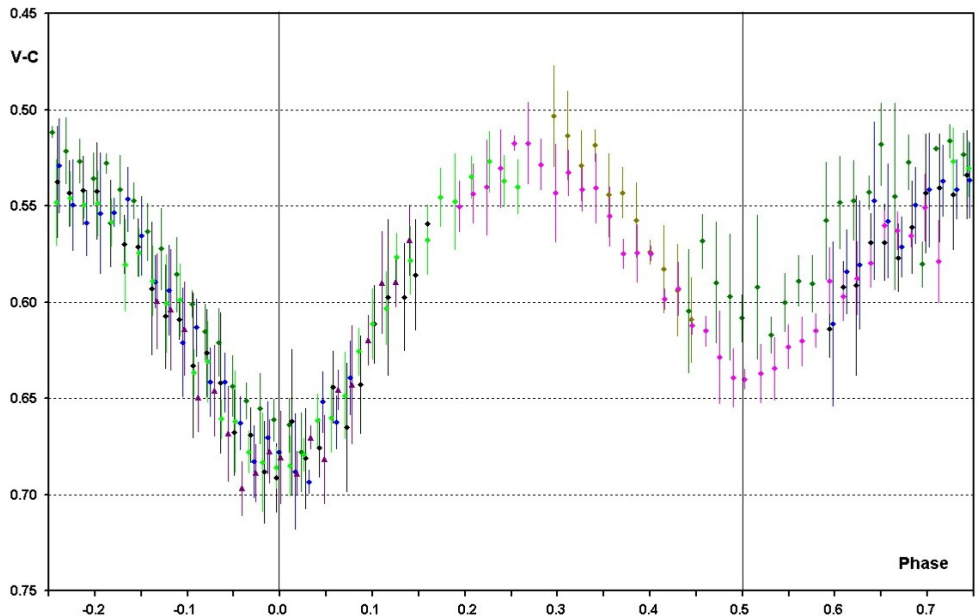

**Figure 12:** The light curve of the star GSC2.3 N25F001920.

The GSC2.3 lists an R magnitude of 14.9 mag. for this star. The faintness of the star leads to an increased scatter, so that the mean of five consecutive measurements was taken. In the LC (fig. 12), the standard deviation for each point has been indicated. The star is probably a short period eclipsing variable(of W Uma type?) with shallow eclipses of 0.15 mag. amplitude for Min. I and 0.1 mag. for Min. II.

Both seven min. I and min. II (table 3) were derived and led to the ephemeris

$$HJD\ (Min.) = 2456064.5155(13) + 0.389228(18) * E \quad (4)$$

| J.D. Hel. |  | Epoch | weight | (O-C) | J.D. Hel. |  | Epoch | weight | (O-C) |
|---|---|---|---|---|---|---|---|---|---|
| 2456064.517 | Min. I | 0.0 | 10 | 0.001 | 2456075.408 | Min. I | 28.0 | 10 | -0.006 |
| 2456065.485 | Min. II | 2.5 | 10 | -0.004 | 2456076.394 | Min. II | 30.5 | 5 | 0.007 |
| 2456067.435 | Min. II | 7.5 | 10 | 0.000 | 2456090.404 | Min. II | 66.5 | 5 | 0.005 |
| 2456069.379 | Min. II | 12.5 | 10 | -0.002 | 2456096.429 | Min. I | 82.0 | 10 | -0.003 |
| 2456071.526 | Min. I | 18.0 | 10 | 0.004 | 2456102.465 | Min. II | 97.5 | 5 | 0.000 |
| 2456072.498 | Min. II | 20.5 | 10 | 0.003 | 2456110.443 | Min. I | 118.0 | 10 | -0.001 |
| 2456073.467 | Min. I | 23.0 | 10 | -0.001 | 2456131.464 | Min. I | 172.0 | 10 | 0.001 |

**Table 3:** Minima of the star GSC2.3 N25F001920.

**3) Var. XX (18:09:41.141 +32:33:46.99),** GSC2.3 N25D013720, 2MASS 18094113+3233468:

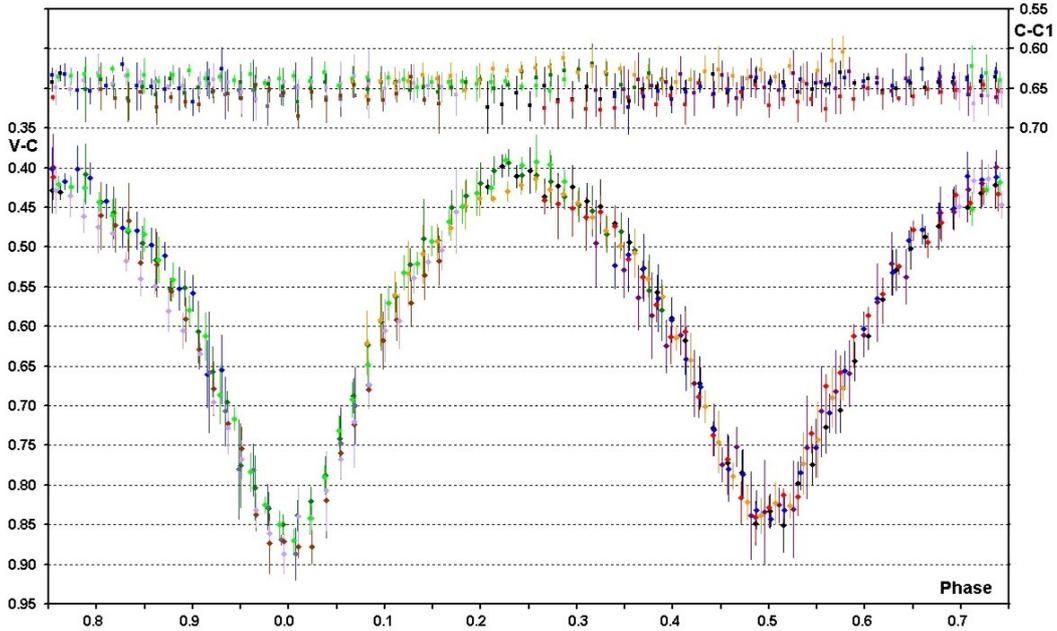

**Figure 13:** The light curve of the star GSC2.3 N25D013720.

A slightly fainter R magnitude of 15.3 mag. is given for this star. Again, five consecutive points were taken as one measurement. The LC (fig. 13) suggests a classical W Uma system with two eclipses of about 0.45 mag. amplitude. Both five min. I and min. II (table 4) were derived and led to the ephemeris:

$$\text{HJD (Min.)} = 2456072.4354(4) + 0.393563(7) * E \quad (5)$$

| J.D. Hel. |  | Epoch | (O-C) | J.D. Hel. |  | Epoch | (O-C) |
|---|---|---|---|---|---|---|---|
| 2456069.4852 | Min II | -7.5 | 0.0015 | 2456087.3919 | Min I | 38.0 | 0.0011 |
| 2456071.4500 | Min II | -2.5 | -0.0015 | 2456096.4437 | Min I | 61.0 | 0.0010 |
| 2456072.4349 | Min I | 0.0 | -0.0005 | 2456102.5421 | Min II | 76.5 | -0.0008 |
| 2456073.4194 | Min II | 2.5 | 0.0001 | 2456105.4943 | Min I | 84.0 | -0.0003 |
| 2456084.4382 | Min II | 30.5 | -0.0008 | 2456131.4699 | Min I | 150.0 | 0.0001 |

**Table 4:** Minima of the star GSC2.3 N25D013720.

**4) Var. XXX (18:09:01.767 +32:36:53.59),** GSC2.3 N25D015132, 2MASS 18090177+3236533:

An even fainter R magnitude of 16.13 mag. is given for this star, which is at the limit of the equipment. But the LC appears sufficiently well defined to suggest that the star is also a W Uma system, this time with a shorter period and an unequal depht of the min. with 0.75 and 0.5 mag. amplitude. Four min. I and six min. II led to the ephemeris:

$$\text{HJD (Min.)} = 245072.4909(2) + 0.287800(3) * E \quad (6)$$

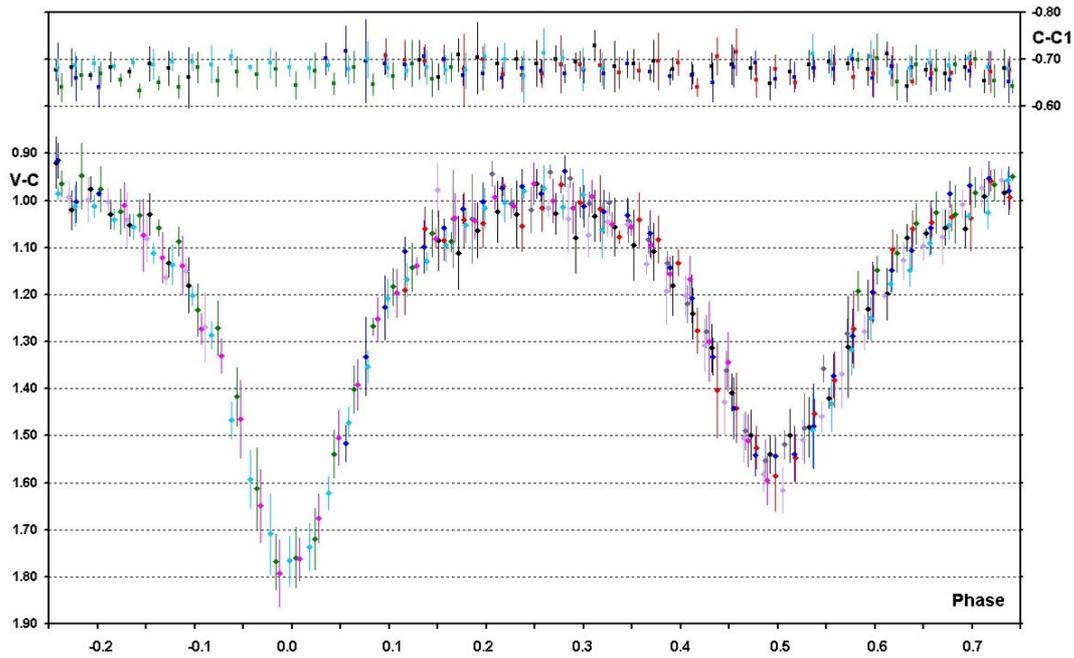

**Figure 14:** The light curve of the star GSC2.3 N25D015132.

| J.D. Hel. |        | Epoch | (O-C)   | J.D. Hel. |        | Epoch | (O-C)   |
|-----------|--------|-------|---------|-----------|--------|-------|---------|
| 2456065.4382 | Min II | -24.5 | -0.0016 | 2456090.4776 | Min II | 62.5  | -0.0008 |
| 2456069.4695 | Min II | -10.5 | 0.0005  | 2456102.4220 | Min I  | 104.0 | -0.0001 |
| 2456071.4834 | Min II | -3.5  | -0.0002 | 2456105.4442 | Min II | 114.5 | 0.0002  |
| 2456072.4912 | Min I  | 0.0   | 0.0003  | 2456117.3878 | Min I  | 156.0 | 0.0001  |
| 2456073.4988 | Min II | 3.5   | 0.0006  | 2456131.4898 | Min I  | 205.0 | -0.0001 |

**Table 5:** Minima of the star GSC2.3 N25D015132.

**Aknowledgements:**

In this research, data from the WASP public archive was used. The WASP consortium comprises of the University of Cambridge, Keele University, University of Leicester, The Open University, The Queen's University Belfast, St. Andrews University and the Isaac Newton Group. It was also made use of the VizieR and Aladin databases operated at the Centre de Données Astronomiques (Strasbourg) in France.
I would like to thank Stefan Hümmerich for improving the manuscript, Klaus Bernhard and Joachim Hübscher for their advices.

**References:**

Butters et al., 2010, SuperWASP Public archive, http://www.wasp.le.ac.uk/public/lc/index.php

Collier Cameron et al., 2006, A fast hybrid algorithm for exoplanetary transit searches, MNRAS, Volume 373, Issue 2, pp. 799-810, arXiv:astro-ph/0609418

Le Borgne, J.F. et al., 2007, The all–Sky GEOS RR Lyr Survey with the TAROT Telescope, Analysis of the Blazhko Effect, arXiv:astro-ph/1205.6397v1

Motl, D., 2012, http://c-munipack.sourceforge.net/

Srdoc, G., Bernhard, K., 2012, Zwei neue RR Lyrae Sterne aus der SuperWasp Datenbank, BAV Rundbrief 61, 1.

The Catalina Surveys, CSDR2, http://nesssi.cacr.caltech.edu/DataRelease/

**Appendices:**

Appendix 1: Maxima and minima of the RR Lyrae star GSC 02626-00896.

Appendix 2: The 2012 measurements of the RR Lyrae star GSC 02626-00896.

**Maxima of the RR Lyrae star GSC 02626-00896:**

| Max. HJD | Epoch | weight | (O-C) |
|---|---|---|---|
| 2453141.650 | -9072 | 5 | -0.014 |
| 2453144.542 | -9063 | 5 | -0.026 |
| 2453150.682 | -9044 | 5 | -0.018 |
| 2453151.651 | -9041 | 5 | -0.017 |
| 2453152.619 | -9038 | 5 | -0.017 |
| 2453153.615 | -9035 | 5 | 0.011 |
| 2453154.585 | -9032 | 5 | 0.013 |
| 2453155.548 | -9029 | 5 | 0.007 |
| 2453156.545 | -9026 | 5 | 0.036 |
| 2453157.514 | -9023 | 5 | 0.037 |
| 2453158.484 | -9020 | 5 | 0.039 |
| 2453162.654 | -9007 | 5 | 0.014 |
| 2453163.617 | -9004 | 5 | 0.008 |
| 2453164.577 | -9001 | 5 | 0.000 |
| 2453165.539 | -8998 | 5 | -0.006 |
| 2453166.503 | -8995 | 5 | -0.010 |
| 2453170.687 | -8982 | 5 | -0.021 |
| 2453171.655 | -8979 | 5 | -0.022 |
| 2453172.624 | -8976 | 5 | -0.021 |
| 2453173.595 | -8973 | 5 | -0.018 |
| 2453174.563 | -8970 | 5 | -0.018 |
| 2453175.527 | -8967 | 5 | -0.022 |
| 2453176.495 | -8964 | 5 | -0.022 |
| 2453178.442 | -8958 | 5 | -0.012 |
| 2453180.430 | -8952 | 5 | 0.040 |
| 2453183.654 | -8942 | 5 | 0.037 |
| 2453184.623 | -8939 | 5 | 0.038 |
| 2453185.586 | -8936 | 5 | 0.032 |
| 2453192.639 | -8914 | 5 | -0.015 |
| 2453194.577 | -8908 | 5 | -0.013 |
| 2453195.539 | -8905 | 5 | -0.019 |
| 2453196.506 | -8902 | 5 | -0.020 |
| 2453197.477 | -8899 | 5 | -0.017 |
| 2453198.440 | -8896 | 5 | -0.022 |
| 2453199.405 | -8893 | 5 | -0.026 |
| 2453203.614 | -8880 | 5 | -0.012 |
| 2453204.578 | -8877 | 5 | -0.016 |
| 2453205.553 | -8874 | 5 | -0.009 |
| 2453207.540 | -8868 | 5 | 0.042 |
| 2453208.498 | -8865 | 5 | 0.031 |
| 2453209.470 | -8862 | 5 | 0.035 |
| 2453215.580 | -8843 | 5 | 0.013 |
| 2453217.509 | -8837 | 5 | 0.006 |
| 2453218.475 | -8834 | 5 | 0.004 |
| 2453219.432 | -8831 | 5 | -0.007 |
| 2453226.523 | -8809 | 5 | -0.016 |
| 2453227.495 | -8806 | 5 | -0.012 |
| 2453228.468 | -8803 | 5 | -0.008 |
| 2453229.428 | -8800 | 5 | -0.016 |
| 2453230.396 | -8797 | 5 | -0.016 |
| 2453238.507 | -8772 | 5 | 0.027 |
| 2453239.470 | -8769 | 5 | 0.022 |
| 2453240.434 | -8766 | 5 | 0.018 |
| 2453241.397 | -8763 | 5 | 0.013 |
| 2453248.463 | -8741 | 5 | -0.021 |
| 2453249.442 | -8738 | 5 | -0.010 |
| 2453250.408 | -8735 | 5 | -0.013 |
| 2453258.470 | -8710 | 5 | -0.019 |
| 2453260.450 | -8704 | 5 | 0.025 |
| 2453261.432 | -8701 | 5 | 0.039 |
| 2453262.396 | -8698 | 5 | 0.035 |
| 2453263.353 | -8695 | 5 | 0.024 |
| 2453271.391 | -8670 | 5 | -0.007 |
| 2453272.353 | -8667 | 5 | -0.013 |
| 2454303.450 | -5472 | 5 | -0.011 |
| 2454304.422 | -5469 | 5 | -0.007 |
| 2454312.502 | -5444 | 5 | 0.005 |
| 2454321.534 | -5416 | 5 | 0.001 |
| 2454322.502 | -5413 | 5 | 0.001 |
| 2454324.433 | -5407 | 5 | -0.004 |
| 2454325.409 | -5404 | 5 | 0.003 |
| 2454332.497 | -5382 | 5 | -0.008 |
| 2454333.475 | -5379 | 5 | 0.001 |
| 2454334.437 | -5376 | 5 | -0.005 |
| 2454335.408 | -5373 | 5 | -0.002 |
| 2454579.706 | -4616 | 5 | -0.004 |
| 2454580.685 | -4613 | 5 | 0.007 |
| 2454591.642 | -4579 | 5 | -0.009 |
| 2454593.580 | -4573 | 5 | -0.007 |
| 2454600.693 | -4551 | 5 | 0.006 |
| 2454605.525 | -4536 | 5 | -0.003 |
| 2454613.598 | -4511 | 5 | 0.002 |
| 2454614.566 | -4508 | 5 | 0.002 |
| 2454621.658 | -4486 | 5 | -0.006 |
| 2454622.632 | -4483 | 5 | 0.000 |
| 2454623.606 | -4480 | 5 | 0.006 |
| 2454624.566 | -4477 | 5 | -0.002 |
| 2454625.537 | -4474 | 5 | 0.001 |
| 2454626.503 | -4471 | 5 | -0.002 |
| 2454627.460 | -4468 | 5 | -0.013 |
| 2454630.687 | -4458 | 5 | -0.013 |
| 2454631.653 | -4455 | 5 | -0.015 |
| 2454632.645 | -4452 | 5 | 0.009 |
| 2454635.541 | -4443 | 5 | 0.000 |
| 2454636.508 | -4440 | 5 | -0.001 |
| 2454637.479 | -4437 | 5 | 0.002 |
| 2454638.448 | -4434 | 5 | 0.003 |
| 2454641.657 | -4424 | 5 | -0.016 |
| 2454642.632 | -4421 | 5 | -0.009 |
| 2454643.609 | -4418 | 5 | 0.000 |
| 2454644.577 | -4415 | 5 | 0.000 |
| 2454645.548 | -4412 | 5 | 0.003 |
| 2454646.509 | -4409 | 5 | -0.004 |
| 2454652.632 | -4390 | 5 | -0.013 |
| 2454655.553 | -4381 | 5 | 0.003 |
| 2454656.516 | -4378 | 5 | -0.002 |
| 2454657.491 | -4375 | 5 | 0.005 |
| 2454665.548 | -4350 | 5 | -0.006 |
| 2454666.527 | -4347 | 5 | 0.005 |
| 2454668.461 | -4341 | 5 | 0.003 |
| 2454675.548 | -4319 | 5 | -0.010 |
| 2454676.527 | -4316 | 5 | 0.001 |
| 2454677.493 | -4313 | 5 | -0.002 |
| 2454688.473 | -4279 | 5 | 0.006 |
| 2456064.556 | -15 | 10 | 0.005 |
| 2456065.524 | -12 | 10 | 0.005 |
| 2456069.404 | 0 | 10 | 0.012 |

**Minima of the RR Lyrae star GSC 02626-00896:**

| Min. JD hel. | Epoch | weight | (O-C) |
|---|---|---|---|
| 2453128.652 | -9097 | 5 | 0.013 |
| 2453129.615 | -9094 | 5 | 0.008 |
| 2453130.597 | -9091 | 5 | 0.022 |
| 2453137.677 | -9069 | 5 | 0.002 |
| 2453138.642 | -9066 | 5 | -0.002 |
| 2453139.615 | -9063 | 5 | 0.003 |
| 2453141.540 | -9057 | 5 | -0.008 |
| 2453150.603 | -9029 | 5 | 0.019 |
| 2453151.554 | -9026 | 5 | 0.002 |
| 2453152.540 | -9023 | 5 | 0.019 |
| 2453153.492 | -9020 | 5 | 0.003 |
| 2453158.635 | -9004 | 5 | -0.017 |
| 2453159.629 | -9001 | 5 | 0.009 |
| 2453160.590 | -8998 | 5 | 0.001 |
| 2453161.572 | -8995 | 5 | 0.015 |
| 2453162.527 | -8992 | 5 | 0.002 |
| 2453163.507 | -8989 | 5 | 0.014 |
| 2453167.691 | -8976 | 5 | 0.003 |
| 2453168.645 | -8973 | 5 | -0.012 |
| 2453169.605 | -8970 | 5 | -0.020 |
| 2453170.582 | -8967 | 5 | -0.011 |
| 2453171.541 | -8964 | 5 | -0.020 |
| 2453172.528 | -8961 | 5 | -0.001 |
| 2453173.515 | -8958 | 5 | 0.018 |
| 2453174.458 | -8955 | 5 | -0.008 |
| 2453177.687 | -8945 | 5 | -0.006 |
| 2453178.665 | -8942 | 5 | 0.004 |
| 2453179.631 | -8939 | 5 | 0.002 |
| 2453180.604 | -8936 | 5 | 0.007 |
| 2453181.570 | -8933 | 5 | 0.005 |
| 2453182.540 | -8930 | 5 | 0.006 |
| 2453183.509 | -8927 | 5 | 0.007 |
| 2453184.493 | -8924 | 5 | 0.023 |
| 2453185.453 | -8921 | 5 | 0.015 |
| 2453190.609 | -8905 | 5 | 0.007 |
| 2453191.564 | -8902 | 5 | -0.006 |
| 2453192.527 | -8899 | 5 | -0.011 |
| 2453193.498 | -8896 | 5 | -0.008 |
| 2453194.474 | -8893 | 5 | 0.000 |
| 2453195.418 | -8890 | 5 | -0.025 |
| 2453200.596 | -8874 | 5 | -0.010 |
| 2453201.576 | -8871 | 5 | 0.002 |
| 2453202.537 | -8868 | 5 | -0.005 |
| 2453203.522 | -8865 | 5 | 0.011 |
| 2453212.562 | -8837 | 5 | 0.015 |
| 2453215.465 | -8828 | 5 | 0.014 |
| 2453222.530 | -8806 | 5 | -0.021 |
| 2453223.503 | -8803 | 5 | -0.016 |
| 2453224.465 | -8800 | 5 | -0.022 |
| 2453225.430 | -8797 | 5 | -0.026 |
| 2453226.427 | -8794 | 5 | 0.003 |
| 2453227.388 | -8791 | 5 | -0.004 |
| 2453235.460 | -8766 | 5 | 0.000 |
| 2453236.435 | -8763 | 5 | 0.007 |
| 2453245.473 | -8735 | 5 | 0.009 |
| 2453246.428 | -8732 | 5 | -0.005 |
| 2453255.486 | -8704 | 5 | 0.017 |
| 2453256.458 | -8701 | 5 | 0.021 |
| 2453258.387 | -8695 | 5 | 0.014 |
| 2454297.535 | -5475 | 5 | -0.002 |
| 2454298.502 | -5472 | 5 | -0.003 |
| 2454306.579 | -5447 | 5 | 0.006 |
| 2454307.547 | -5444 | 5 | 0.006 |
| 2454308.513 | -5441 | 5 | 0.004 |
| 2454318.512 | -5410 | 5 | -0.002 |
| 2454320.449 | -5404 | 5 | -0.001 |
| 2454321.425 | -5401 | 5 | 0.007 |
| 2454328.512 | -5379 | 5 | -0.006 |
| 2454329.484 | -5376 | 5 | -0.002 |
| 2454330.450 | -5373 | 5 | -0.004 |
| 2454331.423 | -5370 | 5 | 0.001 |
| 2454575.717 | -4613 | 5 | -0.006 |
| 2454576.695 | -4610 | 5 | 0.004 |
| 2454577.664 | -4607 | 5 | 0.005 |
| 2454578.620 | -4604 | 5 | -0.007 |
| 2454579.605 | -4601 | 5 | 0.010 |
| 2454585.710 | -4582 | 5 | -0.017 |
| 2454586.684 | -4579 | 5 | -0.011 |
| 2454597.662 | -4545 | 5 | -0.006 |
| 2454598.631 | -4542 | 5 | -0.005 |
| 2454600.568 | -4536 | 5 | -0.004 |
| 2454606.698 | -4517 | 5 | -0.006 |
| 2454607.671 | -4514 | 5 | -0.001 |
| 2454608.646 | -4511 | 5 | 0.006 |
| 2454609.608 | -4508 | 5 | 0.000 |
| 2454618.643 | -4480 | 5 | -0.002 |
| 2454619.608 | -4477 | 5 | -0.005 |
| 2454620.582 | -4474 | 5 | 0.001 |
| 2454621.547 | -4471 | 5 | -0.002 |
| 2454622.518 | -4468 | 5 | 0.001 |
| 2454623.482 | -4465 | 5 | -0.003 |
| 2454627.677 | -4452 | 5 | -0.004 |
| 2454628.643 | -4449 | 5 | -0.006 |
| 2454629.613 | -4446 | 5 | -0.004 |
| 2454630.586 | -4443 | 5 | 0.001 |
| 2454631.551 | -4440 | 5 | -0.002 |
| 2454632.515 | -4437 | 5 | -0.007 |
| 2454638.647 | -4418 | 5 | -0.006 |
| 2454639.620 | -4415 | 5 | -0.002 |
| 2454640.587 | -4412 | 5 | -0.003 |
| 2454641.556 | -4409 | 5 | -0.002 |
| 2454642.521 | -4406 | 5 | -0.005 |
| 2454643.489 | -4403 | 5 | -0.005 |
| 2454649.625 | -4384 | 5 | -0.001 |
| 2454650.587 | -4381 | 5 | -0.007 |
| 2454651.549 | -4378 | 5 | -0.013 |
| 2454652.519 | -4375 | 5 | -0.011 |
| 2454655.425 | -4366 | 5 | -0.010 |
| 2454656.400 | -4363 | 5 | -0.003 |
| 2454660.597 | -4350 | 5 | -0.001 |
| 2454661.559 | -4347 | 5 | -0.008 |
| 2454662.533 | -4344 | 5 | -0.002 |
| 2454663.505 | -4341 | 5 | 0.002 |
| 2454664.473 | -4338 | 5 | 0.002 |
| 2454671.573 | -4316 | 5 | 0.002 |
| 2454672.535 | -4313 | 5 | -0.004 |
| 2454673.512 | -4310 | 5 | 0.005 |
| 2454674.473 | -4307 | 5 | -0.002 |
| 2454675.438 | -4304 | 5 | -0.006 |
| 2454682.529 | -4282 | 5 | -0.014 |
| 2454683.513 | -4279 | 5 | 0.001 |
| 2454684.472 | -4276 | 5 | -0.008 |
| 2454685.456 | -4273 | 5 | 0.008 |
| 2454686.421 | -4270 | 5 | 0.005 |
| 2456064.438 | 0 | 10 | 0.001 |
| 2456065.408 | 3 | 10 | 0.003 |
| 2456067.346 | 9 | 10 | 0.004 |
| 2456071.540 | 22 | 10 | 0.003 |
| 2456072.507 | 25 | 10 | 0.002 |
| 2456073.477 | 28 | 10 | 0.004 |
| 2456075.411 | 34 | 10 | 0.001 |
| 2456076.385 | 37 | 10 | 0.007 |
| 2456084.447 | 62 | 10 | 0.001 |
| 2456096.392 | 99 | 10 | 0.005 |
| 2456102.522 | 118 | 10 | 0.004 |
| 2456105.428 | 127 | 10 | 0.005 |